\documentclass[review]{elsarticle}

\newcommand{\sect}[1]{\setcounter{equation}{0}\section{#1}
\renewcommand{\theequation}{\arabic{section}.\arabic{equation}}}
\newcommand{\subsect}[1]{\setcounter{equation}{0}\subsection{#1}
\renewcommand{\theequation}{\arabic{section}.\arabic{subsection}.\arabic{equation}}}

\usepackage{amssymb}
\usepackage{graphicx}

\journal{Physica A}
\begin{document}

\begin{frontmatter}

\title{$\kappa$-Deformed Fourier Transform}
\author{A.M. Scarfone}
\address{Istituto dei Sistemi Complessi (ISC-CNR) c/o
Politecnico di Torino\\ Corso Duca
degli Abruzzi 24, 10129 Torino, Italy.}
\ead{antoniomaria.scarfone@cnr.it}
\begin{abstract}
We present a new formulation of Fourier transform in the picture of the $\kappa$-algebra derived in the framework of the $\kappa$-generalized statistical mechanics. The $\kappa$-Fourier transform is obtained from a $\kappa$-Fourier series recently introduced by us [2013 Entropy {\bf15} 624]. The kernel of this transform, that reduces to the usual exponential phase in the $\kappa\to0$ limit, is composed by a $\kappa$-deformed phase and a damping factor that gives a wavelet-like behavior. We show that the $\kappa$-Fourier transform is isomorph to the standard Fourier transform through a changing of time and frequency variables. Nevertheless, the new formalism is useful to study, according to Fourier analysis, those functions defined in the realm of the $\kappa$-algebra. As a relevant application, we discuss the central limit theorem for the $\kappa$-sum of $n$-iterate statistically independent random variables.
\end{abstract}


\end{frontmatter}


\sect{Introduction}
As well known, Fourier transform is a very useful tool widely used in mathematical statistics and physics. It consists of a linear operator acting on a function $f(x)$, with a real
argument $x$, and transforming it into a function $\widehat f(\omega)\equiv{\cal F}[f(x)](\omega)$, with a complex argument $\omega$:
\begin{eqnarray}
{\cal F}[f(x)](\omega)={1\over\sqrt{2\,\pi}}\int\limits_{-\infty}\limits^{+\infty}f(x)\,e^{-i\,x\,\omega}\,dx \ .
\end{eqnarray}
More in general, it belongs to the family of integral transforms, defined in
\begin{eqnarray}
{\cal H}[f(x)](\omega)=\int\limits_{-\infty}\limits^{+\infty}f(x)\,h(x,\,\omega)\,dx \ ,\label{standard}
\end{eqnarray}
like Laplace, Mellin and Hankel transforms, with the kernel $h(x,\,\omega)$ given by a complex exponential-like form. \\
Fourier transform finds applications in many fields of science, running from mathematics to physics and engineering. In statistical physics, and in general in probability theory, it is applied in the study of random walk, of infinite divisible distributions, in the analysis of distribution stability and in the study of their domains of attraction as well as in the large deviation theory and in the proof of several central limit theorems \cite{Feller}.\\
In quantum mechanics, Fourier transform links the representation of a wave function in the configuration space with its dual representation in the momenta space. In addition, it is a powerful tool widely used in quantum field theory to evaluate the Green function of a quantum propagator thanks to its property of changing a derivative in a multiplicative power.\\
Its discrete version is largely applied in the theory of signals \cite{Oppenheim}, especially after  the introduction of fast algorithms, the so called fast Fourier transform, that speeds up the computation process, allowing the elaboration of a large amount of information, which has permitted to develop relevant data compression algorithms and the realization of high-definition image devices.\\ In quantum computing theory, Fourier transform is powerfully used in the generation of quantum protocols for the factorization of large numbers and in solving the discrete logarithm problem \cite{Nielsen}.

In the past, several versions of Fourier transform have been proposed in literature by using deformed algebraic structures like, for instance, in the framework of quantum groups where the Hopf algebra underlying the braiding groups  has been used to define a Fourier transform in the noncommutative space \cite{Schirmacher}. In \cite{Coulembier}, within the {\em basic}-calculus, it has been advanced a {\em basic}-analogous of standard Fourier transform defined in
\begin{eqnarray}
{\cal F}[f(x);\,q](\omega)=\int\limits_{-\infty}\limits^{+\infty}f(x)\exp(x|\omega;q)\,d_qx \ ,
\end{eqnarray}
that is realized by employing a complex version of the {\em basic}-exponential $\exp(x;\,q)$ \cite{Exton} and the {\em basic}-integral $\int d_qx$ \cite{Gasper}.\\
On a different ground, in the framework of statistical mechanics based on the Tsallis entropy \cite{Tsallis}, the nonlinear Fourier transform \cite{Umarov} defined in
\begin{eqnarray}
{\cal F}_q[f(x)](\omega)=\int\limits_{-\infty}\limits^{+\infty}f(x)\otimes_q
\exp_q(-i\,x\,\omega)\,dx \ ,\label{qF}
\end{eqnarray}
has been introduced by using the $q$-exponential \cite{Tsallis1} and the $q$-product \cite{Borges}. This transform was designed with the purpose to investigate the possible generalization of the central limit theorem within the $q$-statistics, although the final results are still controversial \cite{Hilhorst,Umarov1}.

In this paper, we revisit the standard Fourier transform in a manner that is consistent with the $\kappa$-algebra and the $\kappa$-calculus derived in the framework of the $\kappa$-statistical mechanics. This formalism, together with the $\kappa$-entropy, has been introduced in \cite{Kaniadakis0} to generalize the Boltzmann-Gibbs entropy and the related theory, with the aim to describe non Gibbsian systems characterized by power-law distributions. In the last decade, several papers have been written on the $\kappa$-statistical mechanics concerning its foundations, its theoretical consistence and its possible applications to physical and physical-like systems \cite{Wada,Wada1,Quarati,Pereira,Scarfone,Trivellato} (see \cite{Kaniadakis1} for an up to date references list).\\
The new formulation of Fourier transform (hereinafter $\kappa$-Fourier transform), is obtained starting from a $\kappa$-deformed version of Fourier series recently proposed in \cite{Scarfone-1}, given by the following linear operator
\begin{eqnarray}
{\cal F}_\kappa[f(x)](\omega)={1\over\sqrt{2\,\pi}}\int\limits_{-\infty}\limits^{+\infty}f(x)\,
\exp_\kappa(-x\otimes_\kappa\omega)^i\,d_\kappa x \ ,\label{new}
\end{eqnarray}
and is based on the $\kappa$-exponential and the related $\kappa$-algebra introduced in \cite{Kaniadakis0} (see also \cite{Kaniadakis}), by replacing both the exponential kernel and the integrate operator with their $\kappa$-deformed versions. Here and in the following, for the sake of notation, $\exp_\kappa(x)^a$ means ($\exp_\kappa(x))^a$. \\
Actually, Eq. (\ref{new}) can be rewritten in the form (\ref{standard}) of a standard integral transform over the real numbers with the kernel
\begin{eqnarray}
h_\kappa(x,\,\omega)={\exp(-i\,x_{\{\kappa\}}\,\omega_{\{\kappa\}})\over\sqrt{1+\kappa^2\,x^2}} \ ,
\end{eqnarray}
where $x_{\{\kappa\}}$ are $\kappa$-numbers defined in the remainder \cite{Kaniadakis}.\\
Function $h_\kappa(x,\,\omega)$ contains a deformed phase factor $\exp(-i\,x_{\{\kappa\}}\,\omega_{\{\kappa\}})$ which has an asymptotically log-periodic behavior and a damping factor $1/\sqrt{1+\kappa^2\,x^2}$, that confers to the transform a wavelet-like behaviours. Furthermore, transform (\ref{new}) is isomorph to the standard Fourier transform, where the isomorphism is settled by changing the variables in time and frequency.

The plan of the paper is as follows. In the next Section 2, we summarize the main aspects of the $\kappa$-algebra and its related calculus. We introduce the Euler formula and the cyclic deformed functions in a manner consistent with the $\kappa$-formalism. In Section 3, we derive the new formulation of the Fourier transform based on the $\kappa$-algebra. A list of its main properties is also given, while a potential application concerning the central limit theorem for $\kappa$-sum of $n$-iterate statistically independent random variables is discussed in Section 4. Our conclusions are reported in Section 5.

\sect{Mathematical formalism of the $\kappa$-algebra}

In order to embody power-law distributions in statistical physics, in \cite{Kaniadakis0} it has been proposed the following generalized entropic form
\begin{eqnarray}
S_\kappa[p]=-\int p(x)\,\ln_\kappa\big(p(x)\big)\,dx \ ,\label{k}
\end{eqnarray}
which mimics the Boltzmann-Gibbs entropy by replacing the standard logarithm with its generalized version
\begin{eqnarray}
\ln_\kappa(x)&=&
\frac{x^{\kappa}-x^{-\kappa}}{2\,\kappa}\equiv{1\over\kappa}\,\sinh\big(\kappa\,\ln(x)\big) \ ,\label{lnk}
\end{eqnarray}
In agreement with the maximal entropy principle, the corresponding equilibrium distribution reads
\begin{eqnarray}
p(x)=\alpha\,\exp_\kappa
\left(-{1\over\lambda}\,(\gamma+\beta\,x)\right)
\ ,\label{dist}
\end{eqnarray}
where
\begin{eqnarray}
\exp_\kappa(x)&=&\left(\kappa\, x+\sqrt{1+\kappa^2\,x^2}\right)^{1/\kappa}\equiv\exp\left({1\over\kappa}\,{\rm
arcsinh}\,(\kappa\,x)\right) \ ,\label{expk}
\end{eqnarray}
is the $\kappa$-exponential, with $\ln_\kappa(\exp_\kappa(x))=\exp_\kappa(\ln_\kappa(x))=x$.\\
Both these functions reduce to the standard exponential and logarithm in the $\kappa\to0$ limit: $\ln_0(x)\equiv\ln(x)$ and $\exp_0(x)\equiv\exp(x)$ and consequently, in the same limit, Eqs. (\ref{k}) and (\ref{dist}) converge to the Boltzmann-Gibbs entropy and the Gibbs distribution, respectively.\\
We note that the shape of distribution (\ref{dist}), is characterized by an asymptotic power law tail, since $\exp_\kappa(x)\sim x^{\pm1/\kappa}$, for $x\to\pm\infty$. Therefore, it differs from the exponential behavior of the Gibbs-distribution, a fact that justify the use of $\kappa$-statistical mechanics in the study of free-scale systems.\\
On a mathematical ground, the main features of $\kappa$-statistical physics follow from the analytical properties of the $\kappa$-exponential and the $\kappa$-logarithm \cite{Kaniadakis}.\\
For any $|\kappa|<1$, $\log_\kappa(x)$ and $\exp_\kappa(x)$ are continuous, monotonic, increasing functions, normalized in $\ln_\kappa(1)=0$ and $\exp_\kappa(0)=1$, with $\ln_\kappa(\mathbb{R}^+)\subseteq \mathbb{R}$ and $\exp_\kappa(\mathbb{R})\subseteq \mathbb{R}^+$.

Another function useful in the formulation of the $\kappa$-deformed statistical mechanics is given by
\begin{eqnarray}
u_\kappa(x)=\frac{x^\kappa+x^{-\kappa}}{2}\equiv{1\over\kappa}\,\cosh\big(\kappa\,\ln(x)\big) \ .\label{u}
\end{eqnarray}
For any $|\kappa|<1$, the function $u_\kappa(x)$ is continuous, with $u_\kappa(\mathbb{R}^+)\in\mathbb{R}^+$, $u_\kappa(0)=u_\kappa(+\infty)=+\infty$
and reaches its minimum at $x=1$, where $u_\kappa(1)=1$. Function $u_\kappa(x)$, that reduces to a constant in the $\kappa\to0$ limit ($u_0(x)=1$), is strictly related to $\ln_\kappa(x)$ and $\exp_\kappa(x)$ and appears recurrently in the study of their analytical properties.\\
These $\kappa$-functions, fulfil the following scaling-laws
\begin{eqnarray}
&&\exp_\kappa(\mu\,x)=\exp_{\kappa'}(x)^\mu \ ,\label{se}\\
&&\ln_\kappa(x^\mu)=\mu\,\ln_{\kappa'}(x) \ ,\\
&&u_\kappa(x^\mu)=u_{\kappa'}(x) \ ,
\end{eqnarray}
where $\kappa'=\mu\,\kappa$. In particular, for $\mu=-1$ and accounting for the symmetry relations $\exp_\kappa(x) \equiv\exp_{-\kappa}(x)$, $\ln_\kappa(x)\equiv\ln_{-\kappa}(x)$ and $u_\kappa(x)\equiv u_{-\kappa}(x)$,
we obtain
\begin{eqnarray}
&&\exp_\kappa(x)\,\exp_\kappa(-x)=1 \ ,\label{eq1}\\
&&\ln_\kappa(x)+\ln_\kappa(1/x)=0 \ ,\label{eq2}\\
&&u_\kappa(x)-u_\kappa(1/x)=0 \ .\label{eq3}
\end{eqnarray}
Equations (\ref{eq1}) and (\ref{eq2}) reproduce the well-known relations of standard exponential and logarithm in the $\kappa$-formalism.

\subsect{$\kappa$-Algebra}

In \cite{Scarfone1}, it has been shown that, starting from a pair of continuous, monotonic increasing functions, inverse each other, we can introduce two different algebraic structures endowed by a generalized sum and product, that form two distinct Abelian fields. By specializing this result to the present case, we can define two deformed algebras on the target space and on the probability space, respectively. Among them, the target space algebra is the relevant one in this work and is revisited in the following, remanding to \cite{Scarfone1} for a more detailed discussion.\\ To begin with, let us introduce the $\kappa$-numbers
\begin{eqnarray}
x_{\{\kappa\}}=\frac{1}{\kappa}\, {\rm arcsinh}
\,(\kappa\,x) \ ,\label{xk1}
\end{eqnarray}
and their dual
\begin{eqnarray}
x^{\{\kappa\}}=\frac{1}{\kappa}\,\sinh
\,(\kappa\,x) \ ,\label{xk2}
\end{eqnarray}
with
\begin{eqnarray}
\left(x_{\{\kappa\}}\right)^{\{\kappa\}}=\left(x^{\{\kappa\}}\right)_{\{\kappa\}}=x \ .\label{dual}
\end{eqnarray}
Mapping (\ref{xk1}) [resp. (\ref{xk2})] defines a non uniform stretching of the real axis so that the space of the $\kappa$-numbers $\mathbb{R}^\kappa$ is isomorph to the space of the real numbers, with $(+\infty)_{\{\kappa\}}\equiv+\infty$, $(-\infty)_{\{\kappa\}}\equiv-\infty$ and $0_{\{\kappa\}}\equiv0$.\\
Generalized sum and product are defined on $\mathbb{R}^\kappa$ according to \cite{Scarfone-1}
\begin{eqnarray}
x\oplus_\kappa y=\left(x_{\{\kappa\}}+ y_{\{\kappa\}}\right)^{\{\kappa\}} \ ,\label{kpiu}\\
x\otimes_\kappa y=\left(x_{\{\kappa\}}\cdot y_{\{\kappa\}}\right)^{\{\kappa\}} \ ,\label{kper}
\end{eqnarray}
where, in the following, for the sake of notation, we simply mean $\oplus\equiv\oplus_\kappa$ and $\otimes\equiv\otimes_\kappa$ unless explicably stated.\\
These operations are associative, commutative and distributed, with the null element $\emptyset$ and the identity $I$, and for any $x\in\mathbb{R}^\kappa$, there exist the opposite $(-x)$ and the inverse $(1/x)$.
Therefore, the algebraic structure
$(\mathbb{R}^\kappa,\,\oplus,\,\otimes)$ forms a commutative group.\\
Explicitly, we have
\begin{eqnarray}
&&x\oplus y={1\over\kappa}\,\sinh\Big(\,{\rm
arcsinh}\,(\kappa\,x)+{\rm arcsinh}\,(\kappa\,y)\,\Big) \
,\label{ksum}\\
&&x\otimes y={1\over\kappa}\,\sinh\left({1\over\kappa}\,{\rm
arcsinh}\,(\kappa\,x)\cdot{\rm arcsinh}\,(\kappa\,y)\right) \
,\label{kprod}
\end{eqnarray}
with $\emptyset\equiv0$, $I\equiv\kappa^{-1}\,\,\sinh\kappa$, $(-x)\equiv-x$ and $(1/x)\equiv\kappa^{-1}\,\sinh(\kappa^2/{\rm arcsinh}\,\kappa\,x)$.\\
In the $\kappa\to0$ limit, sum (\ref{ksum}) and product (\ref{kprod}) recover the standard operations and the $\kappa$-algebra reduces to the standard algebra of the real numbers.\\
Remark also that, for large $x$ and $y$, the $\kappa$-sum approaches asymptotically to the standard product
\begin{eqnarray}
  x\oplus y\simeq x\cdot y \ ,\qquad \ x,\,y\gg1 \ ,\label{okp}
\end{eqnarray}
for $\kappa\not=0$, as well as
\begin{eqnarray}
  x\oplus y\simeq x+y \ ,\qquad \ x,\,y\to0 \ .
\end{eqnarray}
Starting from Eqs. (\ref{ksum}) and (\ref{kprod}), we can introduce other elementary operations like the difference $x\ominus y= x\oplus(-y)$ and the quotient $x\oslash y=x\otimes(1/y)$.\\
However, the main property of $\kappa$-algebra follows from the definitions of $\kappa$-exponential and $\kappa$-logarithm
\begin{eqnarray}
\exp_\kappa(x)=e^{x_{\{\kappa\}}} \ ,\qquad\ln_\kappa(x)=\left(\ln x\right)_{\{\kappa\}} \ .\label{ke}
\end{eqnarray}
Among the many relations satisfied by $\exp_\kappa(x)$ and $\ln_\kappa(x)$, let us recall that
\begin{eqnarray}
&&\exp_\kappa(x\oplus y)=\exp_\kappa(x)\cdot\exp_\kappa(y) \ ,\\
&&\ln_\kappa(x\cdot y)=\ln_\kappa(x)\oplus\ln_\kappa(y) \ ,
\end{eqnarray}
as well as the following, that will be used in the remainder
\begin{eqnarray}
\nonumber
&&\exp_\kappa(x\otimes y)=\exp_\kappa (x)^{y_{\{\kappa\}}}=\exp_\kappa(y)^{x_{\{\kappa\}}} \ ,\\
&&\exp_\kappa(a\,x\otimes_\kappa a\,y)=\exp_{\kappa^\prime}(x\otimes_{\kappa^\prime} y)^{a^2}
=\exp_{\kappa^{\prime\prime}}(a^2(x\otimes_{\kappa^\prime} y)) \ ,\label{eax}\\
\nonumber
&&\exp_\kappa(a\,x\oplus_{\kappa}a\,y)=\exp_{\kappa^\prime}(x\oplus_{\kappa^\prime}y)^{a}=\exp_\kappa
\left(a(x\oplus_{\kappa^\prime} y)\right) \ ,
\end{eqnarray}
with $\kappa^\prime=\kappa\,a$ and $\kappa^{\prime\prime}=\kappa/a$. Other relations can be obtained by inspection of \cite{Scarfone-1}.

\subsect{$\kappa$-Cyclic functions}

According to the Euler formula, the cyclic functions in the $\kappa$-formalism can be introduced starting from the $\kappa$-deformed version of the complex exponential. However,
within the $\kappa$-algebra, we have two substantially different possible definitions \cite{Scarfone-1}. The most natural one is given by
\begin{eqnarray}
\exp_\kappa(x)\to\exp_\kappa(i\,x)\equiv\exp\left((i\,x)_{\{\kappa\}}\right)  \ ,\label{ei1}
\end{eqnarray}
and coincides with the choice made in \cite{Kaniadakis1}. The function $\exp_\kappa(i\,x)$ is unitary for $|x|\leq1/|\kappa|$, while becomes a purely imaginary quantity with increasing modulo for $|x|>1/|\kappa|$, that is
\begin{eqnarray}
&&|\exp_\kappa(i\,x)|=1\hspace{41mm}{\rm for \ }|x|\leq1/|\kappa| \ ,\\
&&|\exp_\kappa(i\,x)|=\left(\kappa\,x+\sqrt{\kappa^2\,x^2-1}\right)^{1/\kappa}\qquad{\rm for \ }|x|>1/|\kappa| \ .
\end{eqnarray}
Consistently, with definition (\ref{ei1}) we can introduce the $\kappa$-trigonometric functions according to
\begin{eqnarray}
  \exp_\kappa(i\,x)={\rm Cos}_\kappa(x)+i\,{\rm Sin}_\kappa(x) \ ,
\end{eqnarray}
so that
\begin{eqnarray}
&&{\rm Sin}_\kappa(x)={\exp_\kappa(i\,x)-\exp_\kappa(-i\,x)\over2\,i} \ ,\\ &&{\rm Cos}_\kappa(x)={\exp_\kappa(i\,x)+\exp_\kappa(-i\,x)\over2} \ .\label{ck}
\end{eqnarray}
These functions take real values in the range $|x|\leq1/|\kappa|$ and imaginary values otherwise. This is illustrated in Figure 1, where we plot the ${\rm Cos}_\kappa(x)$ for $\kappa=0.05$, in the region $|x|\leq1/|\kappa|$ (left panel) and its modulo $|{\rm Cos}_\kappa(x)|$ (right panel) in the region $|x|>1/|\kappa|$, where the $\kappa$-cosine (\ref{ck}) assumes imaginary values.
\begin{figure}[h]
\begin{center}
\includegraphics*[width=11cm]{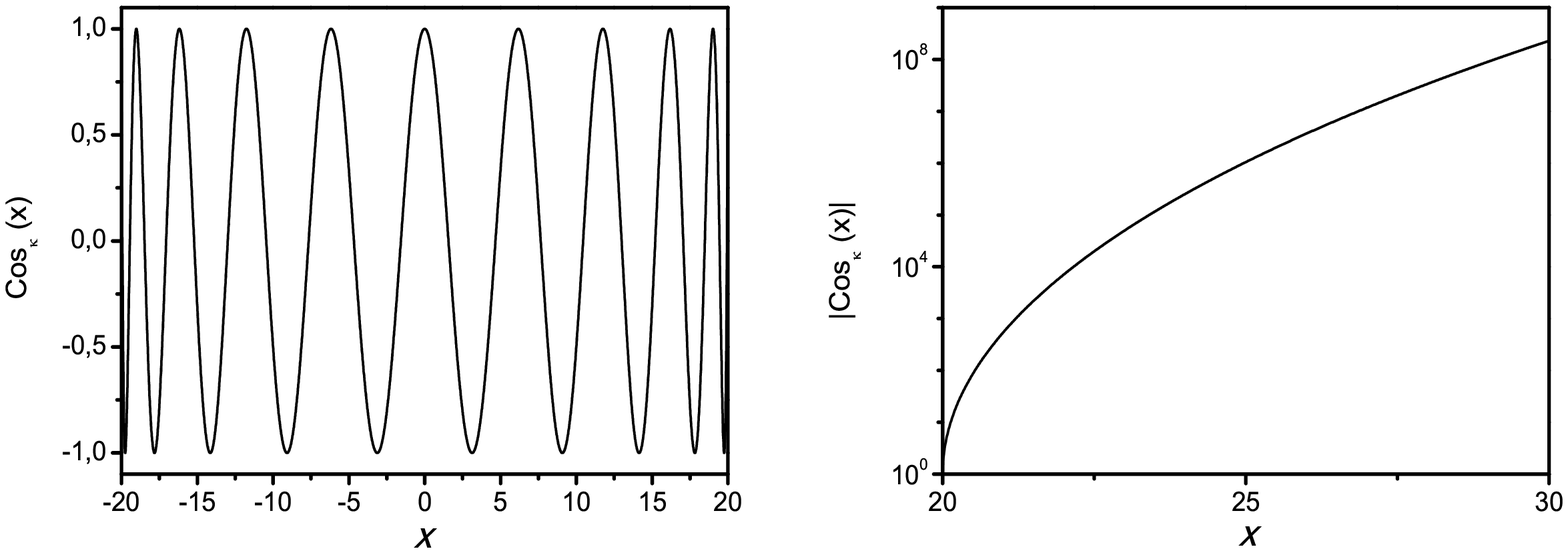}\\
\caption{Plot of ${\rm Cos}_\kappa(x)$ defined in Eq. (\ref{ck}), for $\kappa=0.05$, in the region $|x|\leq 1/|\kappa|$ (left panel) and its modulo in the region $|x|>1/|\kappa|$ (right panel).}
\end{center}
\end{figure}

A different definition of the $\kappa$-exponential on the complex unit circle is given by
\begin{eqnarray}
\exp_\kappa(x)\to\exp_\kappa (x)^i\equiv\exp\left(i\,x_{\{\kappa\}}\right) \ .\label{ei}
\end{eqnarray}
Actually, definitions (\ref{ei1}) and (\ref{ei}) are related each other in
\begin{eqnarray}
\exp_\kappa(x)^i=\exp_{\kappa'}(i\,x) \ ,\label{ei2}
\end{eqnarray}
according to the scaling relation (\ref{se}), with $\kappa'=-i\,\kappa$.\\
However, the complex function $\exp_\kappa(x)^i$ has unitary modulo for any $x\in\mathbb{R}$ and, by taking into account that $|x|<|x_{\{\kappa\}}|$ and $|x-x_{\{\kappa\}}|$ increases for $|x|\to\infty$, it follows that the period of functions $\exp_\kappa(x)^i$ increases as $|x|$ increases.\\
According to Eq. (\ref{ei}), we introduce a new set of $\kappa$-deformed cyclic functions as
\begin{eqnarray}
\exp_\kappa(x)^i=\cos_\kappa(x)+i\,\sin_\kappa(x) \ ,
\end{eqnarray}
so that
\begin{eqnarray}
&&\sin_\kappa(x)\equiv\sin(x_{\{\kappa\}}) \ ,\label{ksin}\\
&&\cos_\kappa(x)\equiv\cos(x_{\{\kappa\}}) \ .\label{kcos}
\end{eqnarray}
\begin{figure}[h]
\begin{center}
\includegraphics*[width=11cm]{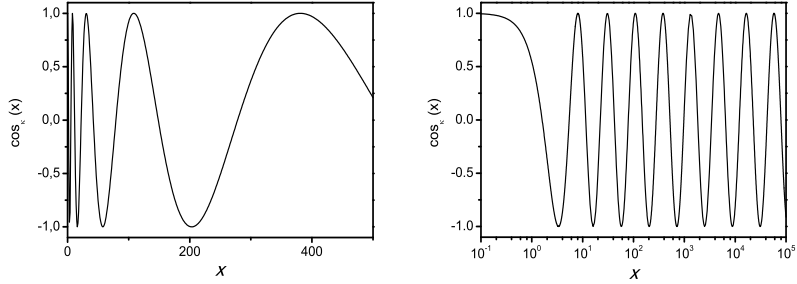}\\
\caption{Plot of $\cos_\kappa(x)$ given by Eq. (\ref{kcos}) for $k=0.2$ in the linear-linear scale (a) and in the log-linear scale (b) showing its asymptotic log-periodic behavior.}
\end{center}
\end{figure}
We remark that functions (\ref{ksin}) and (\ref{kcos}) are asymptotically log-periodic. Their period increases for $|x|\to\infty$, because
\begin{eqnarray}
\sin_\kappa(x) =\sin_\kappa(x^\prime) \ , \qquad{\rm when}\qquad x^\prime=\left(x_{\{\kappa\}}+2\,n\,\pi\right)^{\{\kappa\}} \ ,
\end{eqnarray}
and, for large $x$, we have
\begin{eqnarray}
\Delta\ln(x)\simeq2\,n\,\kappa\,\pi \ ,\qquad{\rm that \ is}\qquad x^\prime\simeq x\,e^{2\,n\,\kappa\,\pi} \ ,\label{period}
\end{eqnarray}
where $\Delta\ln(x)=\ln(x^\prime)-\ln(x)$ approaches to a constant for $x\gg1$.\\
This is shown in Figure 2, where we plot the shape of $\cos_\kappa(x)$, for $\kappa=0.2$, in the linear-linear scale (left panel) and in the log-linear scale (right panel) that shows its asymptotic log-periodic behavior.

As proven in \cite{Scarfone-1}, the functions $\sin_\kappa(x)$ and $\cos_\kappa(x)$ can be derived from the following differential equation
\begin{eqnarray}
{d\over d x}\left(\sqrt{1+\kappa^2\,x^2}\,{d\,u(x)\over dx}\right)+{\left(a_{\{\kappa\}}\right)^2\over\sqrt{1+\kappa^2\,x^2}}\,u(x)=0 \ ,\label{ksl}
\end{eqnarray}
with $-h\leq x\leq h$ and $a_{\{\kappa\}}$ a constant. This is a Sturm-Liouville equation
\begin{eqnarray}
{d\over dx}\left(p(x)\,{df(x)\over dx}\right)+\lambda\,w(x)\,f(x)=0 \ ,
\end{eqnarray}
with $p(x)=\sqrt{1+\kappa^2\,x^2}$, weight function $w(x)=1/\sqrt{1+\kappa^2\,x^2}$ and eigenvalue $\lambda=(a_{\{\kappa\}})^2$. A solution of Eq. (\ref{ksl}), with boundary conditions $f(-h)=f(h)=0$, corresponds to the $\kappa$-sine function
\begin{eqnarray}
f(x)=A\,\sin_\kappa(a\otimes x) \ ,
\end{eqnarray}
whilst a solution with boundary conditions $f^\prime(-h)=f^\prime(h)=0$, where prime means derivative with respect to its argument, is given by the $\kappa$-cosine function
\begin{eqnarray}
f(x)=A\,\cos_\kappa(a\otimes x) \ .
\end{eqnarray}
Similar considerations hold for the function
\begin{eqnarray}
f(x)=A\,\exp_\kappa(a\otimes x)^i \ ,
\end{eqnarray}
that is the solution of the same Sturm-Liouville problem, in the $h\to\infty$ limit, with boundary condition $|f(x)|<\infty$.

\subsect{$\kappa$-Calculus}

Following \cite{Kaniadakis0}, we introduce the calculus in the $\kappa$-formalism by means of the $\kappa$-differential $d_\kappa x\equiv dx_{\{\kappa\}}$, given by the differential of the $\kappa$-numbers
\begin{eqnarray}
d_\kappa x={dx\over\sqrt{1+\kappa^2\,x^2}} \ ,\label{dk}
\end{eqnarray}
that is $\kappa$-linear
\begin{eqnarray}
&&d_\kappa(a\otimes x)=a_{\{\kappa\}}\cdot d_\kappa x \ ,\\
&&d_\kappa(x\oplus y)=d_\kappa x+d_\kappa y \ ,
\end{eqnarray}
where $a$ is constant.\\
The $\kappa$-derivative is defined in
\begin{eqnarray}
\left({d\over dx}\right)_\kappa\equiv {d\over d_\kappa x} \ ,
\end{eqnarray}
and is related to the Leibnitz derivative by
\begin{eqnarray}
\frac{d}{d_\kappa x}=\sqrt{1+\kappa^2\,x^2}\,{d\over dx}\equiv u_\kappa\left(\exp_\kappa(x)\right)\,{d\over dx} \ ,\label{dk1}
\end{eqnarray}
where the last equality follows from the relation $u_\kappa(\exp_\kappa(x))=\sqrt{1+\kappa^2\,x^2}$ derived in \cite{Scarfone4}.

Within the standard calculus, the derivative of the $\kappa$-functions previously introduced read
\begin{eqnarray}
&&{d\over dx}\exp_\kappa(x)={\exp_\kappa(x)\over u_\kappa\left(\exp_\kappa(x)\right)} \ ,\label{de1}\\
&&{d\over dx}\ln_\kappa(x)={u_\kappa(x)\over x} \ ,\label{de2}\\
&&{d\over dx}u_\kappa(x)=\kappa\,{\ln_\kappa(x)\over x} \ ,\\
&&{d\over dx}\sin_\kappa(x)={\cos_\kappa(x)\over u_\kappa\left(\exp_\kappa(x)\right)} \ ,\\
&&{d\over dx}\cos_\kappa(x)=-{\sin_\kappa(x)\over u_\kappa\left(\exp_\kappa(x)\right)} \ .\label{de3}
\end{eqnarray}
In this way, by using Eq. (\ref{dk1}), we can rewrite these relations in
\begin{eqnarray}
&&\frac{d}{d_\kappa x}\,\exp_\kappa(x)=\exp_\kappa(x) \ ,\\
&&\frac{d}{d_\kappa x}\,\sin_\kappa(x)=\cos_\kappa(x) \ ,\\
&&\frac{d}{d_\kappa x}\,\cos_\kappa(x)=-\sin_\kappa(x) \ ,
\end{eqnarray}
that are consistent within the $\kappa$-formalism.\\
We observe that quantities like $\exp_\kappa(x)\,d_\kappa x$, $\sin_\kappa(x)\,d_\kappa x$ and  $\cos_\kappa(x)\,d_\kappa x$ are all exact differentials in the standard calculus, since they correspond, respectively to the following differential forms: $d\exp_\kappa(x)$, $d\sin_\kappa(x)$ and $d\cos_\kappa(x)$.\\
In addition, accounting for the $\kappa$-linearity, we can verify the relations
\begin{eqnarray}
&&\frac{d}{d_\kappa x}f(x\oplus y)=\frac{d f(z)}{d_\kappa z}\Big|_{z=x\oplus y} \ ,\\
&&\frac{d}{d_\kappa x}\exp_\kappa(x\otimes y)=y_{\{\kappa\}}\frac{d f(z)}{d_\kappa z}\Big|_{z=x\otimes y} \ ,
\end{eqnarray}
and in particular
\begin{eqnarray}
&&\frac{d}{d_\kappa x}\exp_\kappa(x\oplus y)=\exp_\kappa (x\oplus y) \ ,\\
&&\frac{d}{d_\kappa x}\,\exp_\kappa(x\otimes y)=y_{\{\kappa\}}\,\exp_\kappa(x\otimes y) \ ,
\end{eqnarray}
that will be used in the remainder.

Finally, let us introduce the $\kappa$-integral $\int f(x)\,d_\kappa x$, as the inverse operator to the $\kappa$-derivative, according to
\begin{eqnarray}
\left({d\over d x}\right)_\kappa\left(\int f(x)\,d_\kappa x+const.\right)=f(x) \ ,
\end{eqnarray}
extending, in this way, the fundamental integral theorem to the $\kappa$-formalism.\\
We observe that the $\kappa$-integral can be written like a weighted ordinary integral. In fact, recalling Eq. (\ref{dk}), we have
\begin{eqnarray}
\int f(x)\,d_\kappa x=\int f(x)\,w(x)\,dx\equiv\int {f(x)\over\sqrt{1+\kappa^2\,x^2}}\,dx \ ,\label{int1}
\end{eqnarray}
where the weight function
\begin{eqnarray}
w(x)={1\over u_\kappa\left(\exp_\kappa(x)\right)}\equiv{1\over\sqrt{1+\kappa^2\,x^2}} \ ,
\end{eqnarray}
coincides with that introduced in the Sturm-Liouville problem.\\
In addition, we can also use the following relation
\begin{eqnarray}
  \int\limits_a\limits^bf(x)\,d_\kappa x=\int\limits_{a_{\{\kappa\}}}\limits^{b_{\{\kappa\}}}f^{\{\kappa\}}(x)\,dx \ ,
\end{eqnarray}
where
\begin{eqnarray}
f^{\{\kappa\}}(x)\equiv f(x^{\{\kappa\}})=f\left({1\over\kappa}\sinh(\kappa\,x)\right) \ ,
\end{eqnarray}
that links the $\kappa$-integral on the $\kappa$-numbers to a standard integral on the real numbers.

\sect{$\kappa$-Fourier transform}

\subsect{The transform}

In \cite{Scarfone-1}, it has been introduced a consistent formulation of Fourier series in the framework of the $\kappa$-algebra. There, it has been shown as any square-integrable function $f(x):\,(-h,\,h)\to\mathbb{R}$, may be expanded in the $\kappa$-Fourier series with respect to the orthogonal and complete system of functions $\sin_\kappa(a_n\otimes x)$ and $\cos_\kappa(a_n\otimes x)$, that is
\begin{eqnarray}
  f(x)=c_0+\sum_{n=1}^\infty \left(s_n\,\sin_\kappa(a_n\otimes x)+c_n\,\cos_\kappa(a_n\otimes x)\right) \ ,\label{kfs}
\end{eqnarray}
where $a_n=(n\,\pi/h_{\{\kappa\}})^{\{\kappa\}}$ are suitable constants.\\
This series  expansion can be written in the complex form
\begin{eqnarray}
f(x)=\sum_{n=-\infty}^\infty\gamma_n\,\exp_\kappa(-a_n\otimes x)^i \ ,
\end{eqnarray}
where the complex Fourier coefficients are given by
\begin{eqnarray}
\gamma_n={\sqrt{2\over h}}\int\limits_{-h}\limits^hf(x)\,\exp_\kappa(a_n\otimes x)^i\,d_\kappa x \ ,\label{coef}
\end{eqnarray}
and are related to the real Fourier coefficients in $\gamma_n=(c_n-i\,s_n)/2$ and $\gamma_{-n}=\gamma_n^\ast$.\\
Starting from this result, $\kappa$-deformed Fourier transform can be formally obtained in the $h\to\infty$ limit, as usually done in the standard theory.\\
Therefore, from Eq. (\ref{coef}) we derive the following integral transform for a real function $f(x)$ given by
\begin{eqnarray}
{\cal F}_\kappa[f(x)](\omega)={1\over\sqrt{2\,\pi}}\int\limits_{-\infty}\limits^{+\infty} f(x)\,
\exp_\kappa (-x\otimes\omega)^i\,d_\kappa x \ .\label{kfourier}
\end{eqnarray}
Accounting for Eqs. (\ref{kper}), (\ref{ke}) and (\ref{dk}), this formula can be written like a standard integral transform [cfr. Eq. (\ref{standard})], with the kernel
\begin{eqnarray}
h_\kappa(x,\,\omega)={\exp(-i\,x_{\{\kappa\}}\,\omega_{\{\kappa\}})\over\sqrt{1+\kappa^2x^2}} \ .\label{kkernel}
\end{eqnarray}
In the $\kappa\to0$ limit, kernel $h_\kappa(x,\,\omega)$ reduces to the usual exponential phase $h(x,\,\omega)=\exp(i\,x\,\omega)$ and the $\kappa$-Fourier transform collapses into the standard Fourier transform.

Kernel (\ref{kkernel}) is formed by a deformed phase factor, that arises from the generalization of the complex exponential in its $\kappa$-deformed version and by a damping factor, that arises from the measure of the $\kappa$-integral. As a consequence, the new transform acquires interesting analytical features. For instance, both the real part and the imaginary part of the phase factor have an asymptotically log-periodic behaviour, a fact that may be relevant in the study of log-oscillating phenomena \cite{Scarfone1}. Otherwise, the damping factor confers to the kernel a wavelet-like behaviour, as shown in Figure 3, where we have plotted both the real and the imaginary part of the kernel $h_\kappa(x,\,\omega)$.\\
However, many properties of the standard Fourier transform are preserved in the generalized version provided that they are reformulated in the $\kappa$-formalism.\\
\begin{figure}[h]
\begin{center}
\includegraphics*[width=11cm]{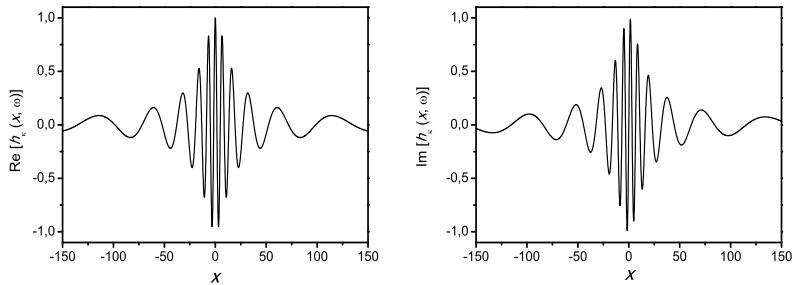}\\
\caption{Real (left panel) and imaginary (right panel) part of the kernel $h_\kappa(x,\,\omega)$, for $k=0.1$ and $\omega=1$.}
\end{center}
\end{figure}

It is duty to note that, in addition to Eq. (\ref{kkernel}), there are other possible generalizations of the kernel. They should satisfy the following reasonable conditions: 1) the kernel must reduce to the standard complex exponential $\exp(i\,\omega\,x)$ in a suitable limit; 2) the kernel should be an unimodular function on the whole real axis. These requests exclude the function
\begin{eqnarray}
h^{(0)}_\kappa(x,\,\omega)=\exp_\kappa(-i\,x\,\omega) \ ,
\end{eqnarray}
since condition 2) is violated in the far region of the real axis, $|x\,\omega|\geq1/|\kappa|$. Otherwise, the following functions are suitable candidates
\begin{eqnarray}
&&h_\kappa^{(1)}(x,\,\omega)=\exp_\kappa (-x)^{i\,\omega} \ ,\label{h1}\\
&&h_\kappa^{(2)}(x,\,\omega)=\exp_\kappa(-\omega)^{i\,x} \ ,\\
&&h_\kappa^{(3)}(x,\,\omega)=\exp_\kappa(-x\,\omega)^i \ ,\label{h3}
\end{eqnarray}
since they have all unitary modulus on $\mathbb{R}$ and converge to $\exp(i\,\omega\,x)$ in the $\kappa\to0$ limit. However, although some of these functions are equivalent each others, like, for instance
\begin{eqnarray}
h_{\kappa}\left(x,\,\omega\right)=h_\kappa^{(1)}\left(x,\,\omega_{\{\kappa\}}\right)
=h_\kappa^{(2)}\left(x_{\{\kappa\}},\,\omega\right) \ ,
\end{eqnarray}
they define different integral transforms with different analytical properties. By inspection, definition (\ref{kfourier}), with the kernel (\ref{kkernel}) is the most consistent choice. In fact, it turns out to be isomorph to the standard transform, so that all the analytical properties of the Fourier transform are preserved when rewritten in the $\kappa$-formalism.\\
Isomorphism follows from a changing of time and frequency variables according to $X=x_{\{\kappa\}}$ and $\Omega=\omega_{\{\kappa\}}$, so that transform (\ref{kfourier}) can be related to a standard Fourier transform of a deformed function
\begin{eqnarray}
{\cal F}_\kappa[f(x)](\omega)\equiv{\cal F}[ f^{\{\kappa\}}(X)]\left(\Omega\right)={1\over\sqrt{2\,\pi}}\int\limits_{-\infty}\limits^{+\infty} f^{\{\kappa\}}(X)\,e^{-i\,X\,\Omega}\,dX \ .\label{kfourier1}
\end{eqnarray}

It is worthy to note that, if the Fourier transform of a function $f(x)$ exists, then certainly exists its $\kappa$-Fourier transform since
\begin{eqnarray}
\Big|{\cal F}_\kappa[f(x)](\omega)\Big|\leq\int\limits_{-\infty}\limits^{+\infty} |f(x)|\,d_\kappa x=\int\limits_{-\infty}\limits^{+\infty} \left|{f(x)\over\sqrt{1+\kappa^2x^2}}\right|\,d x\leq\|f(x)\| \ ,\label{conv}
\end{eqnarray}
where the norm of $f(x)$ is defined in the Banach space $L^1(\mathbb{R})$ as usual
\begin{eqnarray}
\|f(x)\|=\int\limits_{-\infty}\limits^{+\infty}|f(x)|\,dx \ .
\end{eqnarray}
Therefore, the functions space of convergent $\kappa$-Fourier transform is wider than the functions space of convergent standard Fourier transform, thanks to the factor $\sqrt{1+\kappa^2x^2}$ that enforces the convergence of the integral.

Finally, when the function $f(x)$ has a well defined parity, Eq. (\ref{kfourier}) can be rewritten in the form of a $\kappa$-cosine transform
\begin{eqnarray}
{\cal F}_\kappa[f(x)](\omega)={1\over\sqrt{2\,\pi}}\int\limits_0\limits^\infty f(x)\,\cos_\kappa (\omega\otimes x)\,d_\kappa x \ ,
\end{eqnarray}
for even functions or a $\kappa$-sine transform
\begin{eqnarray}
{\cal F}_\kappa[f(x)](\omega)={i\over\sqrt{2\,\pi}}\int\limits_0\limits^\infty f(x)\,\sin_\kappa (\omega\otimes x)\,d_\kappa x \ ,
\end{eqnarray}
for odd functions.

For the sake of illustration, we present in Table 1 the $\kappa$-Fourier transform of several functions. Clearly, the new formulation here proposed has its main advantage in the analysis of $\kappa$-deformed functions. Thus, certain $\kappa$-deformed functions, that can be hardly handled with the standard transform, are easily solved in a closed form with the present formalism, and vice versa. For this reason, we have considered in Table 1 the transform of some deformed functions like $\exp_\kappa(x)$ or $\cos_\kappa(x)$ instead of the corresponding un-deformed versions that, otherwise, are well processed with the standard analysis.\\

\begin{center}
Table 1. $\kappa$-Fourier transform of several simple functions.\\

\begin{tabular}{|c|c|c|}
\hline\hline
&$f(x)$ & ${\cal F}_\kappa[f(x)](\omega)$ \\
\hline\hline
Step function & $\theta(x)$ & $\sqrt{2\,\pi}\,\delta(\omega)+{1\over\sqrt{2\,\pi}\,i\,\omega_{\{\kappa\}}}$ \\
\hline
Modulation & $\cos_\kappa(a\otimes x)$ & $\sqrt{\pi\over2}\,u_\kappa(\exp_\kappa(a))\,\left(\delta(\omega+a)+\delta(\omega-a)\right)$\\
\hline
Causal $\kappa$-exponential & $\theta(x)\,\exp_\kappa(-a\otimes x)$ & ${1\over\sqrt{2\,\pi}}{1\over a_{\{\kappa\}}+i\,\omega_{\{\kappa\}}}$\\
\hline
Symmetric $\kappa$-exponential & $\exp_\kappa(-a\otimes |x|)$ & $ \sqrt{2\over\pi}\,{a_{\{\kappa\}}\over a_{\{\kappa\}}^2+\omega_{\{\kappa\}}^2}$\\
\hline
Constant & $1$ & $\sqrt{2\,\pi}\,\delta(\omega)$\\
\hline
$\kappa$-Phasor & $\exp_\kappa\,(a\otimes x)^i$ & $\sqrt{2\,\pi}\,u_\kappa(\exp_\kappa(a))\,\delta(\omega-a)$\\
\hline
Impuslse & $\delta(x-a)$ & ${1\over\sqrt{2\,\pi}}{\exp_\kappa\,(\omega\otimes a)^i\over u_\kappa\left(\exp_\kappa\,(a)\right)}$\\
\hline
Signum & Sgn$(x)$ & $\sqrt{2\over\pi}\,\,{1\over i\,\omega_{\{\kappa\}}}$\\
\hline
Rectangular & $\Pi\left({x\over a}\right)$ & $\sqrt{2\over\pi}\,\,a_{\{\kappa\}}\,{\rm sinc}_\kappa(\omega\otimes a)$\\
\hline\hline
\end{tabular}
\end{center}

\hspace{10mm}

As an example, we have plotted in Figure 4, the shape of the sinc$_\kappa(x)$ function, corresponding to the $\kappa$-Fourier transform of the rectangular function, for several values of the deformation parameter $\kappa$. The stretching effect produced by the $\kappa$-deformation is clearly observable.
\begin{figure}[h]
\begin{center}
\includegraphics*[width=11cm]{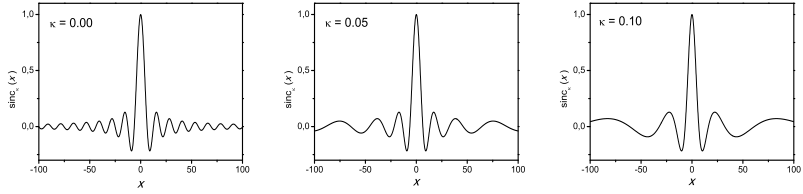}\\
\caption{$\kappa$-sinc function for several values of the deformation parameter $\kappa$.}
\end{center}
\end{figure}

\subsect{The inverse transform}

As expected, the original function $f(x)$ can be reconstructed starting from the transformed function $\widehat f_\kappa(\omega)$ by means of the inverse integral transform, given by
\begin{eqnarray}
{\cal F}^{-1}_\kappa\left[\widehat f_\kappa(\omega)\right](x)={1\over\sqrt{2\,\pi}}\int\limits_{-\infty}\limits^{+\infty}
\widehat f_\kappa(\omega)\,\exp_\kappa (\omega\otimes x)^i\,d_\kappa\omega \ .\label{antikfourier}
\end{eqnarray}
This can be shown by using standard arguments, as follows
\begin{eqnarray}
\nonumber
{\cal F}^{-1}_\kappa\left[{\cal F}_\kappa[f(x)]\right]&=&{1\over2\,\pi}
\int\limits_{-\infty}\limits^{+\infty}
f(y)\,\exp_\kappa (\omega\otimes y)^i\,\exp_\kappa (-\omega\otimes x)^i\,d_\kappa\omega\,d_\kappa y\\
\nonumber
&=&{1\over2\,\pi}\int\limits_{-\infty}\limits^\infty
f(y)\,e^{i\,\omega_{\{\kappa\}}\,\left(y_{\{\kappa\}}-x_{\{\kappa\}}\right)}\,d\omega_{\{\kappa\}}
\,dy_{\{\kappa\}}\\
&=&\int\limits_{-\infty}\limits^\infty
f(y)\,\delta\left(y_{\{\kappa\}}-x_{\{\kappa\}}\right)\,dy_{\{\kappa\}}=f(x)
\ ,\label{proof1}
\end{eqnarray}
where we used the relations
\begin{eqnarray}
\delta\left(y_{\{\kappa\}}-x_{\{\kappa\}}\right)=\delta(y-x)\,\sqrt{1+\kappa^2\,x^2} \ ,\label{kdelta}
\end{eqnarray}
and
\begin{eqnarray}
\delta\left(y_{\{\kappa\}}-x_{\{\kappa\}}\right)\,d_\kappa x=\delta(y-x)\,dx \ .
\end{eqnarray}
Otherwise, plugging Eq. (\ref{antikfourier}) in Eq. (\ref{kfourier}), we obtain
\begin{eqnarray}
\nonumber
{\cal F}_\kappa\left[{\cal F}^{-1}_\kappa[\widehat f_\kappa(\omega)]\right]&=&{1\over2\,\pi}\int\limits_{-\infty}\limits^{+\infty}\widehat f_\kappa(\omega^\prime)\exp_\kappa(-\omega^\prime\otimes x)^i\exp_\kappa(\omega\otimes x)^i\,d_\kappa\omega^\prime\,d_\kappa x\\
\nonumber &=&{1\over2\,\pi}\,\int\limits_{-\infty}\limits^{+\infty}
\widehat f_\kappa(\omega)\,e^{-i\,(\omega_{\{\kappa\}}^\prime-\omega_{\{\kappa\}})\,x_{\{\kappa\}}}\,
d\omega_{\{\kappa\}}^\prime\,dx_{\{\kappa\}}\\
&=&\int\limits_{-\infty}\limits^{+\infty}\widehat f_\kappa(\omega^\prime)\,\delta(\omega_{\{\kappa\}}^\prime-\omega_{\{\kappa\}})\,d\omega_{\{\kappa\}}^\prime=\widehat f_\kappa(\omega) \ .
\end{eqnarray}

Many properties of standard Fourier transforms can be reformulated in the $\kappa$-formalism. For instance, let us consider the $\kappa$-version of the multiplication theorem
\begin{eqnarray}
\int_{-\infty}\limits^{+\infty}f(x)\,g^\ast(x)\,d_\kappa x
=\int_{-\infty}\limits^{+\infty}\widehat f_\kappa(\omega)\,\widehat g_\kappa^{\,\ast}(\omega)\,d_\kappa\omega \ ,\label{multitheo}
\end{eqnarray}
that, rewritten in
\begin{eqnarray}
\int_{-\infty}\limits^{+\infty}{f(x)\,g^\ast(x)\over\sqrt{1+\kappa^2\,x^2}}
\,dx =\int_{-\infty}\limits^{+\infty}{\widehat f_\kappa(\omega)\,\widehat g_\kappa^{\,\ast}(\omega)\over\sqrt{1+\kappa^2\,\omega^2}}\,d\omega \
,
\end{eqnarray}
states a relation between the product of the functions $f(x)$ and $g(x)$ with the product of their $\kappa$-deformed Fourier transforms $\widehat f_\kappa(\omega)$ and $\widehat g_\kappa(\omega)$. Equation (\ref{multitheo}) follows from standard arguments, according to
\begin{eqnarray}
\nonumber
\int_{-\infty}\limits^{+\infty}\widehat f_\kappa(\omega)\widehat g_\kappa^{\,\ast}(\omega)\,d_\kappa\omega&=&\int_{-\infty}\limits^{+\infty}f(x)\,
\exp_\kappa(-\omega\otimes x)^i\,\widehat g^{\,\,\ast}(\omega)\,d_\kappa x\,d_\kappa\omega\\
\nonumber
&=&\int_{-\infty}\limits^{+\infty}f(x)\,\widehat g_\kappa^{\,\ast}(\omega)\,
\exp_\kappa(\omega\otimes x)^i\,d_\kappa\omega\,d_\kappa x\\
&=&\int_{-\infty}\limits^{+\infty}f(x)\,g^\ast(x)\,d_\kappa x \ ,
\end{eqnarray}
and, in particular, for $f(x)=g(x)$, it gives the Plancherel theorem in the $\kappa$-formalism
\begin{eqnarray}
\int_{-\infty}\limits^{+\infty}|f(x)|^2\,d_\kappa x=\int_{-\infty}
\limits^{+\infty}|\widehat f_\kappa(\omega)|^2\,d_\kappa\omega \ .\label{kplanch}
\end{eqnarray}
This relation can be rewritten by means of standard integrals in
\begin{eqnarray}
\int_{-\infty}\limits^{+\infty}{|f(x)|^2\over\sqrt{1+\kappa^2\,x^2}}\,dx=\int_{-\infty}
\limits^{+\infty}{|\widehat f_\kappa(\omega)|^2\over\sqrt{1+\kappa^2\,\omega^2}}\,d\omega \
\end{eqnarray}
that differs clearly from the well known Plancherel theorem formulation. Actually, Eq. (\ref{kplanch}) states a relationships between $\kappa$-Fourier transformed functions and corresponds to the Plancherel relation only in the $\kappa\to0$ limit.

\subsect{Main properties}

Since transform (\ref{kfourier}) can be mapped into a standard Fourier transform, it is not a surprise, has already stated, that many properties of Fourier transform still hold in the deformed version if opportunely rephrased in the $\kappa$-formalism. This is shown in Table 2, where we report the most relevant relations, leaving the reader to verify easily.

\newpage
\begin{center}
Table 2. Main $\kappa$-Fourier properties.\\

\begin{tabular}{|l|l|}

\hline\hline
Linearity &
\small ${\cal F}_\kappa[\alpha\,f(x)+\beta\,g(x)](\omega)=\alpha\,{\cal F}_\kappa[f(x)](\omega)+\beta\,{\cal F}_\kappa[g(x)](\omega)$\\
\hline
Scaling &
\small ${\cal F}_\kappa\left[f(\alpha\,x)\right](\omega)={1\over\alpha}\,{\cal F}_{\kappa^\prime}\left[f(x)\right](\omega^\prime)$\\
 &\small where $\kappa^\prime=\kappa/\alpha$ and $\omega^\prime=(a/\kappa)\,\sinh\left({\rm arcsinh}(\kappa\,\omega)/a^2\right)$\\
\hline
$\kappa$-Scaling &
\small ${\cal F}_\kappa\left[f(\alpha\otimes x)\right](\omega)={1\over\alpha_{\{\kappa\}}}\,{\cal F}_\kappa[f(x)]\left(\frac{1}{\alpha}\otimes\omega\right)$\\
\hline
Complex conjugation &
\small ${\cal F}_\kappa\big[f(x)\big]^{\ast}(\omega)={\cal F}_\kappa\big[f(x)\big](-\omega)$\\
\hline
Duality &
\small ${\cal F}_\kappa\Big[{\cal F}_\kappa\big[f(x)\big](\nu)\Big](\omega)=f(-\omega)$\\
\hline
Reverse &
\small ${\cal F}_\kappa\left[f(-x)\right](\omega)={\cal F}_\kappa[f(x)](-\omega)$\\
\hline
$\kappa$-Frequency shift &
\small ${\cal F}_\kappa\left[\exp_\kappa
(\omega_0\otimes x)^if(x)\right](\omega)={\cal F}_\kappa[f(x)](\omega\ominus\omega_0)$\\
\hline
$\kappa$-Time shift &
\small ${\cal F}_\kappa\left[f(x\oplus x_0)\right](\omega)=\exp_\kappa (\omega\otimes x_0)^i\, {\cal F}_\kappa[f(x)](\omega)$\\
\hline
Transform of $\kappa$-derivative &
\small ${\cal F}_\kappa\left[\frac{d\,f(x)}{d_\kappa x}\right](\omega)=i\,\omega_{\{\kappa\}}\,{\cal F}_\kappa[f(x)](\omega)$\\
\hline
$\kappa$-Derivative of transform &
\small $\frac{d}{d_\kappa\omega}\,{\cal F}_\kappa[f(x)](\omega)=-i\,\omega_{\{\kappa\}}\,{\cal
F}_\kappa\left[x_{\{\kappa\}}\,f(x)\right](\omega)$\\
\hline
Transform of integral &
\small ${\cal F}_\kappa\left[\int\limits_{-\infty}\limits^x
f(y)\,dy\right](\omega)={1\over i\,\omega_{\{\kappa\}}}{\cal F}_\kappa[f(x)](\omega)+2\,\pi\,{\cal F}_\kappa[f(x)](0)\,\delta(\omega)$\\
\hline
$\kappa$-Convolution &
\small ${\cal F}_\kappa\left[(f\mbox{$\bigcirc\hspace{-3mm}*\hspace{1.5mm}$}g)(x)\right](\omega)=\sqrt{2\,\pi}\,{\cal F}_\kappa[f(x)](\omega)\,{\cal F}_\kappa[g(x)](\omega)$\\
 &\small  where $
(f\mbox{$\bigcirc\hspace{-3mm}*\hspace{1.5mm}$}g)(x)=\int\limits_{-\infty}\limits^{+\infty}
f(y)\,g(x\ominus y)\,d_\kappa y $\\
\hline
Modulation &
\small ${\cal F}_\kappa\left[f(x)\,g(x)\right](\omega)={1\over\sqrt{2\,\pi}}\left({\cal F}_\kappa\left[f(x)\right]\mbox{$\bigcirc\hspace{-3mm}*\hspace{1.5mm}$}{\cal F}_\kappa\left[g(x)\right]\right)(\omega)$\\
\hline
\end{tabular}
\end{center}

\sect{Limiting distribution of $\kappa$-sum of $n$-iterate statistically independent random variables}

As known, Fourier transform has many interesting applications both in mathematical statistics and in physics. Among them, the problem of searching for stable distributions of a large number of iterates of independent random variables can be easily handled within the canonical Fourier transform theory. Differently, the problem of searching for stable distributions of a large number of iterates of random variables with a specific statistics dependence or statistically independent random variables with a specific iteration, has been a topic of investigation in the recent years. It can be studied by introducing opportunely defined Fourier transforms \cite{Umarov}.

In this section we are looking for stable distributions of $\kappa$-sum of $n$-iterate statistically independent random variables.\\
In order to introduce this question in the framework of $\kappa$-statistics, let us consider a possible generalization of the characteristic function
\begin{eqnarray}
\varphi\left(f(x),\,\omega\right)=\langle h(x,\,\omega)\rangle\equiv\sqrt{2\,\pi}\,{\cal F}[f(x)](\omega) \ .\label{char}
\end{eqnarray}
For a given normalized probability distribution $f(x)$ of a random variable $X$, is natural to define, in the picture of $\kappa$-algebra, the quantity
\begin{eqnarray}
  \varphi_\kappa\left(f(x),\,\omega\right)=\sqrt{2\,\pi}\,{\cal F}_\kappa[f(x)](\omega) \ ,\label{kchar}
\end{eqnarray}
that mimics Eq. (\ref{char}) recovered in the $\kappa\to0$ limit.\\ Like in the standard case, Eq. (\ref{kchar}) coincides with the linear average of the kernel (\ref{kkernel}) in the space of the real numbers
\begin{eqnarray}
  \varphi_\kappa\left(f(x),\,\omega\right)=\langle h_\kappa(x,\,\omega)\rangle\equiv\int f(x)\,h_\kappa(x,\,\omega)\,dx \ .\label{linear}
\end{eqnarray}
Equivalently, $\varphi_\kappa\left(f(x),\,\omega\right)$ can be written as the $\kappa$-average of the deformed phase factor
\begin{eqnarray}
  \varphi_\kappa\left(f(x),\,\omega\right)={\cal N}\,\langle \exp_\kappa(-x\otimes\omega)^i\rangle_\kappa \ ,\label{fk}
\end{eqnarray}
where the $\kappa$-average is defined in
\begin{eqnarray}
\langle{\cal O}(x)\rangle_\kappa={1\over\cal N}\int{\cal O}(x)\,f(x)\,d_\kappa x \ ,
\end{eqnarray}
with ${\cal N}=\int f(x)\,d_\kappa x$ the normalization constant. Both Eqs. (\ref{linear}) and (\ref{fk}) reduce to the standard characteristic function $\varphi(f(x),\,\omega)=\langle\exp(-i\,x\,\omega)\rangle$ in the $\kappa\to0$ limit.

As an example, let us consider the deformed Gibbs distribution
\begin{eqnarray}
f_\kappa^{\rm G}(x)=\theta(x)\,\exp_\kappa(-\beta\,x) \ , \label{kgibbs}
\end{eqnarray}
where characteristic function $\varphi_{\kappa^\prime}\left(f(x),\,\omega\right)$, with deformation parameter $\kappa^\prime=\beta\,\kappa$, can be easily calculated in
\begin{eqnarray}
\varphi_{\kappa'}\left(f^{\rm G}_\kappa(x),\,\omega\right)={\kappa^\prime\over\kappa^\prime\,\beta-i\,{\rm arcsinh}{(\kappa^\prime\,\omega)}} \ ,
\end{eqnarray}
and recovers, in the $\kappa\to0$ limit, the well-known result $\varphi\left(f^{\rm G}(x),\,\omega\right)=1/(\beta-i\,\omega)$.\\
Otherwise, the standard characteristic function of distribution (\ref{kgibbs}), $\varphi\left(f^{\rm G}_\kappa(x),\,\omega\right)$, assumes a rather complicate expression by means of special functions which are somewhat cumbersome in the analytical manipulation. Remark that, $\varphi_\kappa\left(f(x),\,\omega\right)$ carries similar information of $\varphi\left(f(x),\,\omega\right)$ since both these functions are related to the starting density $f(x)$ by an analytic mapping. In fact, as like as $f(x)$ and $\varphi\left(f(x),\,\omega\right)$ are Fourier transform each other, $f(x)$ and $\varphi_\kappa\left(f(x),\,\omega\right)$ are $\kappa$-Fourier transforms each other. This observation justify the use of the $\kappa$-Fourier transform instead of the classical Fourier transform in the analysis based on the $\kappa$-formalism.\\
Following standard arguments, the phase factor in Eq. (\ref{fk}) can be expanded in powers of $\omega_{\{\kappa\}}$, such that
\begin{eqnarray}
\varphi_\kappa\left(f(x),\,\omega\right)={\cal N}\,\sum_{n=0}^\infty{(-i\,\omega_{\{\kappa\}})^n\over n!}\,\left\langle\left(x_{\{\kappa\}}\right)^n\right\rangle_\kappa \ .\label{series}
\end{eqnarray}
Taking into account the inequality
\begin{eqnarray}
{\cal N}\,\Big\langle\left(x_{\{\kappa\}}\right)^n\Big\rangle_\kappa=\int \left(x_{\{\kappa\}}\right)^n\,f(x)\,d_\kappa x<\int x^n\,f(x)\,dx=\langle x^n\rangle_{\kappa=0} \ ,
\end{eqnarray}
it follows that the $n$-order of the $\kappa$-linear momentum certainly exists if the standard momentum of the same order exists.\\
On the other hand, it is easy to find distributions with only the few first finite standard momenta in spite of the $\kappa$-linear momenta that exist for any order. Again, distribution (\ref{kgibbs}) is a paradigm. In fact, in this case, it is easy to verify that only the few standard momenta with $n<1/\kappa-1$ exist \cite{Scarfone3}, while the $\kappa$-momenta are finite for any order.\\
Finally, from Eq. (\ref{series}) we can obtain the relation
\begin{eqnarray}
\Big\langle\left(x_{\{\kappa\}}\right)^n\Big\rangle_\kappa
={i^n\over\cal N}\,{d^n\varphi_\kappa\left(f(x),\,\omega\right)\over d_\kappa\omega^n}\Bigg|_{\omega=0} \ .
\end{eqnarray}
provided that the left hand side exist, a relation that can be also derived straightforward by using the $\kappa$-derived of transform given in Table 2.

Definition (\ref{kchar}) is useful to study the limit distribution of the $\kappa$-sum of $n$ statistically independent random variables, that is, the limit distribution $f(S_n)$, for large $n$, of $\kappa$-summed random variables
\begin{eqnarray}
  S_n=X_1\oplus X_2\oplus \ldots\oplus X_n \ ,\label{sn}
\end{eqnarray}
given by
\begin{eqnarray}
  f\left(S_n\right)=\Pi_{i=1}^nf(x_i) \ .
\end{eqnarray}
In this case, function $\varphi\left(f(S_n),\,\omega\right)$ coincides with the product of the characteristic functions of $f(x_i)$, i.e.
\begin{eqnarray}
\varphi_\kappa\left(f(S_n),\,\omega\right)=\varphi_\kappa\left(f(x_1),\,\omega\right)
\cdot\varphi_\kappa\left(f(x_2),\,\omega\right)\cdot
\ldots\varphi_\kappa\left(f(x_n),\,\omega\right) \ ,
\end{eqnarray}
as it follows from the distributive property of $\kappa$-sum and $\kappa$-product.
In particular, if the quantities $x_i$ are also identically distributed, then $\varphi_\kappa\left(f(S_n),\,\omega\right)=\varphi_\kappa\left(f(x),\,\omega\right)^n$.

To study the problem of searching families of stable distributions of $\kappa$-sum of $n$-iterate of statistically independent and identically distributed (iid) random variables, we consider a pair of iid random variables $X$, with density distribution $f_1(x)$. The density distribution $f_2(x)$ of $S_2=X\oplus X$, can be obtain from the relation
\begin{eqnarray}
  f_2(y)=\int f_1(x)\,f_1(y\ominus x)\,dx \ .\label{conv1}
\end{eqnarray}
Stable distributions fulfill the condition $f_1(x)=f_2(x)=\ldots=f_n(x)$, where $f_i(x)$, with $i=1,\ldots,\,n$, refers to the pdf of the $i$-iterate $S_i$ given in Eq. (\ref{sn}).\\
They can be derived easily by using the property of $\kappa$-Fourier transform under $\kappa$-convolution (cfr. Table 2), since the characteristic function of stable distributions is invariant under $\kappa$-convolution.\\ Let us consider the following ansatz
\begin{eqnarray}
f_\kappa(x;\,\sigma)=C_\kappa\,\exp_\kappa\left(-{x\over\sqrt{2}\,\sigma}\otimes_\kappa
{x\over\sqrt{2}\,\sigma}\right) \ ,\label{kgauss}
\end{eqnarray}
where $C_\kappa=e^{-\kappa^2/4}/\sqrt{2\,\pi}\,\sigma$ is the normalization constant. Equation (\ref{kgauss}) represents a possible $\kappa$-generalization of Gaussian distribution. Standard Gaussian is recovered in the $\kappa\to0$ limit.\\
Note that, Eq. (\ref{kgauss}) differs from other versions of $\kappa$-Gaussian proposed in the literature \cite{Wada1,Trivellato}. For instance, the following function
\begin{eqnarray}
\widetilde f_\kappa(x)=A_\kappa\,\exp_\kappa\left(-{x^2\over2\,\sigma^2}\right) \ ,\label{kgauss1}
\end{eqnarray}
that corresponds to the asymptotic solution of a diffusive process studied in \cite{Wada1}, has a power law tail different from that of distribution (\ref{kgauss}) that rather decays with a log-normal tail, being
\begin{eqnarray}
f_\kappa(x;\,\sigma)\approx\exp\left(-{1\over\kappa^2}\,\ln^2\left({\sqrt{2}\,\kappa\over\sigma}\,x\right)\right) \ ,
\end{eqnarray}
for $x\gg1$.\\
However, in spite of Eq. (\ref{kgauss1}), Eq. (\ref{kgauss}) is invariant under $\kappa$-Fourier transform, since its $\kappa$-characteristic function $\varphi_{\kappa^\prime}\left(f_\kappa(x;\,\sigma),\,\omega\right)$, with $\kappa^\prime=\kappa/(\sqrt{2}\,\sigma)$, is yet a $\kappa$-Gaussian given by
\begin{eqnarray}
\nonumber
\varphi_{\kappa^\prime}\left(f_\kappa(x;\,\sigma),\,\omega\right)&=&\sigma\, C_\kappa\,\exp_{\kappa^{\prime\prime}}\left(-{\sigma\over\sqrt{2}}\,\omega\otimes_{\kappa^{\prime\prime}}
{\sigma\over\sqrt{2}}\,\omega\right)\\
&\equiv&\sigma\,f_{\kappa^{\prime\prime}}(\omega;\,1/\sigma) \ ,\label{tkg}
\end{eqnarray}
where $\kappa^{\prime\prime}=\kappa/\sigma^2$.\\
More important, function (\ref{kgauss}) is invariant after $n$-iterates of $\kappa$-convolution, defined in
\begin{eqnarray}
  (f\,\mbox{$\bigcirc\hspace{-3.8mm}*\hspace{.5mm}_\kappa$}\,g)(x)=\int\limits_{-\infty}\limits^{+\infty}
f(y)\,g(x\ominus_\kappa y)\,d_\kappa y \ ,\label{conv2}
\end{eqnarray}
that is commutative $(f\,\mbox{$\bigcirc\hspace{-3.8mm}*\hspace{.5mm}_\kappa$}\,g)(x)=(g\,\mbox{$\bigcirc\hspace{-3.8mm}*
\hspace{.5mm}_\kappa$}\,f)(x)$, associative $((f\,\mbox{$\bigcirc\hspace{-3.8mm}*\hspace{.5mm}_\kappa$}\,g)\,\mbox{$\bigcirc\hspace{-3.8mm}*
\hspace{.5mm}_\kappa$}\,h)(x)=
(f\,\mbox{$\bigcirc\hspace{-3.8mm}*\hspace{.5mm}_\kappa$}\,(g\,\mbox{$\bigcirc\hspace{-3.8mm}*
\hspace{.5mm}_\kappa$}\,h))(x)$ and bilinear $((c_1\,f+c_2\,g)\,\mbox{$\bigcirc\hspace{-3.8mm}*\hspace{.5mm}_\kappa$}\,h)(x)=
c_1\,(f\,\mbox{$\bigcirc\hspace{-3.8mm}*\hspace{.5mm}_\kappa$}\,h)
+c_2\,(g\,\mbox{$\bigcirc\hspace{-3.8mm}*\hspace{.5mm}_\kappa$}\,h)$.\\
In fact, posing
\begin{eqnarray}
  F^{(n)}(x;\,\sigma)=(f_\kappa\,\mbox{$\bigcirc\hspace{-3.8mm}*\hspace{.5mm}_{\kappa^\prime}$}\, f_\kappa\,\mbox{$\bigcirc\hspace{-4mm}*\hspace{.5mm}_{\kappa^\prime}$}
  \ldots\mbox{$\bigcirc\hspace{-3.8mm}*\hspace{.5mm}_{\kappa^\prime}$}\,f_\kappa)(x;\,\sigma) \ ,\label{kco}
\end{eqnarray}
its $\kappa$-Fourier transform is related to Eq. (\ref{tkg}) by
\begin{eqnarray}
\varphi_{\kappa^\prime}\left(F^{(n)}(x;\,\sigma),\,\omega\right)=
\left(\varphi_{\kappa^\prime}\left(f_\kappa(x;\,\sigma),\,\omega\right)\right)^n \ .\label{fg}
\end{eqnarray}
On the other hand, accounting for relations (\ref{eax}), the $n$-power of Eq. (\ref{tkg}) can be written in
\begin{eqnarray}
\left(\varphi_{\kappa^\prime}\left(f_\kappa(x;\,\sigma),\,\omega\right)\right)^n=\sigma_n\, C_{\kappa_n}\,\exp_{\kappa^{\prime\prime}_n}\left(-{\sigma_n\over\sqrt{2}}\,\omega\otimes_{\kappa_n^{\prime\prime}}
{\sigma_n\over\sqrt{2}}\,\omega\right) \ ,
\end{eqnarray}
where $\sigma_n=\sigma\,\sqrt{n}$ and $\kappa_n=k\,\sqrt{n}$,
that corresponds to the characteristic of the function
\begin{eqnarray}
f^{(n)}_{\kappa_n}(x,\,\sigma_n)=C_{\kappa_n}\,\exp_{\kappa_n}\left(-{x\over\sqrt{2}\,\sigma_n}\otimes_{\kappa_n}
{x\over\sqrt{2}\,\sigma_n}\right) \ .\label{klc}
\end{eqnarray}

Remark that, the structure of Eq. (\ref{conv2}) used in this proof differs from Eq. (\ref{conv1}) since, the former, contains a $\kappa$-integral instead of a standard integral. Nevertheless, accounting for relation (\ref{int1}), it is easy to verify that the function
\begin{eqnarray}
f_\kappa(x,\,\sigma)={1\over\sqrt{2\,\pi}\,\sigma}{\exp_\kappa\left(-{x\over\sqrt{2}\,\sigma}\otimes_\kappa
{x\over\sqrt{2}\,\sigma}\right)\over\sqrt{1+\kappa^2\,\left(x\over\sqrt{2}\,\sigma\right)^2}} \ ,\label{kg1}
\end{eqnarray}
properly normalized, is stable under composition (\ref{conv1}). Therefore, we can confirm that
\begin{eqnarray}
f^{(n)}_{\kappa_n}(x,\,\sigma_n)={1\over\sqrt{2\,\pi}\,\sigma_n}{\exp_{\kappa_n}\left(-{x\over\sqrt{2}\,\sigma_n}
\otimes_{\kappa_n}
{x\over\sqrt{2}\,\sigma_n}\right)\over\sqrt{1+\kappa_n^2\,\left(x\over\sqrt{2}\,\sigma_n\right)^2}} \ ,\label{kg2}
\end{eqnarray}
is the density distribution of a random variable corresponding to the $\kappa$-sum of $n$-iterate random variables independent and identically distributed according to Eq. (\ref{kg1}).\\
In analogy with the log-normal distribution and the sinh-normal distribution, Eq. (\ref{kg1}) defines a family of arcsinh-normal distributions parameterized by the deformation parameter $\kappa$ and belongs to the family of the Johnson $S_U$ distributions introduced in \cite{Johnson} and given by
\begin{eqnarray}
f(x)={\delta\over\sqrt{2\,\pi}\,\lambda}\,{e^{-{1\over2}\,\left(\gamma+\delta\,{\rm arcsinh}\left({x-\xi\over\lambda}\right)\right)^2}\over\sqrt{1+\left({x-\xi\over\lambda}\right)^2}} \ .\label{Jo}
\end{eqnarray}
Family (\ref{Jo}) coincides with distribution (\ref{kg1}) for $\gamma=\xi=0$, $\delta=\sqrt{2}/\kappa_n$ and $\lambda=\sqrt{2}\,\sigma_n/\kappa_n$.\\
In \cite{Kaniadakis0}  it has been shown that the $\kappa$-sum is substantially equivalent to the relativistic addition of momenta and there it was conjectured a possible relation between the $\kappa$-statistics and the theory of the special relativity. In this sense, the $\kappa$ parameter plays the role of a speed limit according to the relation $\kappa\propto1/c$. Consequently, in the $\kappa\to0$ limit, corresponding to the Galilean relativistic limit ($c\to\infty$), the $\kappa$-sum reduces to the standard sum and consistently, the arcsinh-normal distribution (\ref{kg2}) recovers the Gaussian distribution that, as stated by the standard central limit theorem, is the stable limiting distribution of the sum of iid random variables.

It is fair to note that a similar result has been derived recently in \cite{Keague} by using a different approach.
\begin{figure}[h]
\begin{center}
\includegraphics*[width=11cm]{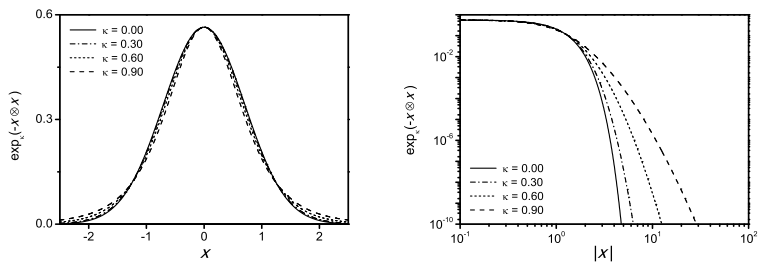}\\
\caption{Plot of $\kappa$-Gaussian (\ref{kg1}) in the linear-linear scale (left panel) and in the log-linear scale (right panel) for several values of $\kappa$. The full-line coincides with the standard Gaussian function.}
\end{center}
\end{figure}
\begin{figure}[h]
\begin{center}
\includegraphics*[width=11cm]{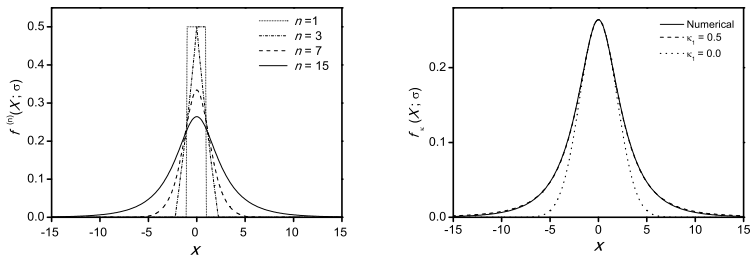}\\
\caption{$\kappa$-convolution of rectangular-function after several iterates, with $\kappa_1=0.5$ (left panel). The comparison of the numerical curve after $n=15$ iterates (full line) with the standard Gaussian (dotted line) and the $\kappa$-Gaussian (\ref{kg2}) (dashed line), are depicted in the right panel.}
\end{center}
\end{figure}
We observe also that, like distribution (\ref{kgauss}), distribution (\ref{kg2}) has a log-normal asymptotic behaviour rather than a power-law tail. This is shown in Figure 5 (left panel), where we have depicted, for the sake of illustration, the shape of the $\kappa$-Gaussian (\ref{kg1}) for several values of the deformation parameter. In the right panel of Figure 5, the same curves are reported in a log-linear scale, showing the log-normal asymptotic behaviour.\\
Finally, like as like the standard Gaussian is the limiting distribution of iid summed random variables, we expect the same holds for distribution (\ref{kg2}) when iid random variables are $\kappa$-summed.
We show the reliability of such statement by means of a numerical computation reported in Figure 6 (left panel) where we plot the distribution of the random variable $S_n$, with $\kappa_1=0.5$, after several $n$-iterations. The starting distribution for the single random variable is assumed to be a rectangular-function. As expected, iterated distribution quickly approaches to a bell-shape. In the same Figure 6 (right panel), this limiting distribution, obtained by a numerical computation after $n=15$ iterates (full line), is compared with the standard Gaussian ($\kappa_1=0$, dotted line) and the $\kappa$-Gaussian (\ref{kg2}) ($\kappa_1=0.5$, dashed line). It is evident the good fit between the numerical curve and the $\kappa$-Gaussian with respect to the standard Gaussian.\\
A further consistency of this result is supported by recalling that the $\kappa$-sum for large values reduces to a standard product [cfr. Eq. (\ref{okp})]. This means that, in the far region of large $x$ values, the random variable $S_n$ corresponds to the product of $n$ iid random variables that, according to the central limit theorem, has a log-normal limit distribution, in agreement with the tail of Eq. (\ref{kg2}).

\sect{Conclusions}

In this paper we have reformulated the standard Fourier transform in a formalism consistent with the $\kappa$-algebra and the $\kappa$-calculus. The new formulation has been derived starting from a $\kappa$-deformed Fourier series recently introduced by us in \cite{Scarfone-1}.\\
The $\kappa$-Fourier transform $\widehat f_\kappa(\omega)\equiv{\cal F}_\kappa[f(x)](\omega)$, belongs to the integral transforms (\ref{standard}) and is characterized by a kernel $h_\kappa(x,\,\omega)$ composed by a deformed phase and a damping factor that confer to $h_\kappa(x,\,\omega)$ a wavelet-like shape. In addition, both the real part and the imaginary part of the phase factor have an asymptotical log-periodic behaviour.\\
We have shown that the $\kappa$-deformed transform of a function $f(x)$ is isomorph to the canonical transform since $\widehat f_\kappa(\omega)$ is equivalent to the standard Fourier transform of the function $f^{\{\kappa\}}(x)\equiv f(x^{\{\kappa\}})$, that is
\begin{eqnarray}
  {\cal F}_\kappa[f(x)](\omega)\equiv{\cal F}[f^{\{\kappa\}}(x)](\omega_{\{\kappa\}}) \ .
\end{eqnarray}
However, in spite of this equivalence, the $\kappa$-Fourier transform turns out to be more appropriate to handle functions defined in the realm of the $\kappa$-algebra.\\
As a relevant example, we have applied our formalism to the study of the limit distribution of $\kappa$-summed statistically independent random variables, that is the distribution given in Eq. (\ref{kg2}) of the random variables $S_n=X\oplus X\oplus \ldots\oplus X$ by using similar arguments employed in the derivation of the central limit theorem. \\

\noindent{\bf References}\\

\vfill\eject
\end{document}